\documentclass[12pt]{article}
\usepackage{epsfig}

\parskip=\medskipamount 
plus.3em minus.5em \abovedisplayshortskip=.5em plus.2em minus.4em
\renewcommand{\thesection}{\arabic{section}}

\catcode`@=11

\@addtoreset{equation}{section} \@addtoreset{equation}{subsection}
\def\theequation{\ifnum\value{section}=0 \arabic{equation}\ignorespaces
\else \ifnum\value{section}=-1 A.\arabic{equation}\ignorespaces
\else \ifnum\value{subsection}=0
\thesection.\arabic{equation}\ignorespaces \else
\thesection.\arabic{subsection}.\arabic{equation}\ignorespaces
                             \fi
                        \fi
                   \fi}

{\catcode`\'=\active \def'{{}^\bgroup\prim@s}}

\catcode`@=12

\newcommand{\bq}{\begin{equation}}
\newcommand{\be}{\begin{equation}}
\newcommand{\fq}{\end{equation}}
\newcommand{\ee}{\end{equation}}
\newcommand{\bqr}{\begin{eqnarray}}
\newcommand{\beqs}{\begin{eqnarray}}
\newcommand{\fqr}{\end{eqnarray}}
\newcommand{\eeqs}{\end{eqnarray}}

\newcommand{\rf}[1]{(\ref{#1})}

\def\b{\beta}

\def\bop#1{\setbox0=\hbox{$#1M$}\mkern1.5mu
    \vbox{\hrule height0pt depth.04\ht0
    \hbox{\vrule width.04\ht0 height.9\ht0 \kern.9\ht0
    \vrule width.04\ht0}\hrule height.04\ht0}\mkern1.5mu}

\begin{document}
\thispagestyle{empty}

\begin{flushright}
\begin{tabular}{l}
hep-th/0503212 \\
\end{tabular}
\end{flushright}

\vskip .6in
\begin{center}

{\Large\bf  Polytopes and Knots}

\vskip .6in

{\bf Gordon Chalmers}
\\[5mm]

{e-mail: gordon@quartz.shango.com}

\vskip .5in minus .2in

{\bf Abstract}

\end{center}

A construction of polytopes is given based on integers.  These 
geometries are constructed through a mapping to pure numbers and 
have multiple applications, including statistical mechanics and 
computer science.  The number form is useful in topology and has 
a mapping to one-dimensional knot contours.  

\setcounter{page}{0}
\newpage
\setcounter{footnote}{0}

\section{Introduction}

Geometric discrete surfaces are commonplace in physics and mathematics.  
Their use in path integrals on discrete spaces, e.g. in statistical 
mechanics or Regge calculus, is well known.  In mathematics   
discrete surfaces are sometimes used to characterize topologies, and 
they have many applied uses.  

The construction of multi-dimensional surfaces in terms of simplicial 
complexes is standard practice in labeling surfaces \cite{Nakahara}.  
Simpicial complexes are not usually written in a convenient form for 
practical computations.  A definition of a simplicial complex, or polytope, 
that is in one to one correspondence with integers is provided in this work.  
The numbering of the surfaces is useful for calculations in mathematics and 
has applications in applied physics including statistical physics.   

A primary example of the use of the polytopic definition presented here 
is many body discrete systems.  Statistical mechanical models require the 
summation of surfaces, i.e. polytopes, weighted in a manner with the coupling 
constants.  The number theoretic definition of the polytopes reduces the many 
body problem of summing the individual lattice sites to one variable.  
Then, the counting of the zeros (e.g. \cite{ChalmersZC}) of an 
associated polynomial generate the solution of statistical mechanics models 
in various dimensions at low temperature, in a well-defined expansion.  

Another important use of the number definition of the polytopes is in 
the construction of alternative computing languages based on geometric 
surfaces.  The gluing and assembly of solids, in a real object oriented 
sense, is relevant to handling of data, but in a number represented form.  
The encoding of data in a geometrically high dimensional sense is also 
useful for information theory and cryptography, and for transcendental 
calculations in mathematics.

\section{Polytope construction: Spatial}

The polytopes considered are labeled by taking a lattice and inserting  
$0$s and $1$s in all of its points.  The $1$s then label a surface.  
Further colorings, e.g. a fiber on the tangent space, on the surface are 
obtained by expanding the base $2$ to base $M$.  This surface is 
illustrated in figure 1.

The polytopes (simplicial complexes) considered are constructed via
a set of integers that label the points and faces parameterizing the
surface.  The integers may be given a matrix representation that
permits a polynomial interpretation, and hence maps to knot(s)
invariant(s).

\begin{figure}
\begin{center}
\epsfxsize=12cm
\epsfysize=12cm
\epsfbox{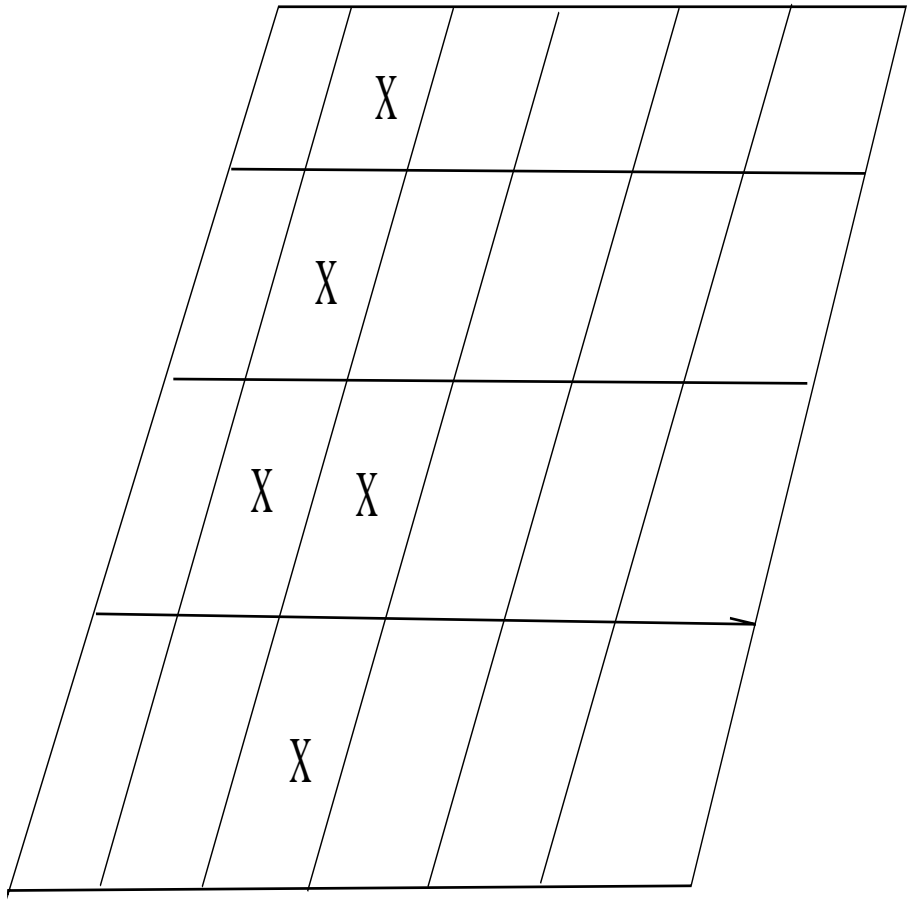}
\end{center}
\caption{An illustration of a polytope (e.g. simplicial complex) in 
two dimensions.}
\end{figure}

The polytopes considered are rectangular in the lattice.  That is the 
solids have edges at right angles in all dimensions.  A straightforward 
generalization alleviates this condition to permit non right angles, 
either by a rotation or a different definition of the polytopic surface 
(and volume).  Also, 'boundary' surfaces, i.e. polytopes constructed with 
only $2$-dimensional surfaces can be defined, generated with a different 
algorithm than used for the space filling surfaces.  

Take a series of numbers $a_1 a_2 \ldots a_n$ corresponding to the
digits of an integer $p$, with the base of the individual number being
$2^n$; this number $a_j$ could be written in base $10$ by the usual
digits.  In this way, upon reduction to base $2$ the digits of the
base reduced number spans a square with $n+1$ entries.  Each number
$a_j$ parameterizes a column with ones and zeros in it.  The lift of
the numbers could be taken to base $10$ with minor modifications,
by converting the base of $p$ to $10$ (with possible remainder issues
if the number does not 'fit' well).

The individual numbers $a_i$ decompose as $\sum a_i^m 2^m$ with
the components $a_i^m$ being $0$ or $1$.  Then map the individual
number to a point on the plane,

\bqr
{\vec r}_i^m  = a_i^m \times m {\hat e}_1 +
       a_i^m \times i {\hat e}_2 \ ,
\fqr
with the original number mapping to a set of points on the plane via
all of the entries in $a_1 a_2 \ldots a_m$.  In doing this, a collection
of points on the plane is spanned by the original number $p$, which
could be a base $10$ number.  The breakdown of the number to a set of
points in the plane is represented in figure 1.

In the case of a rectangular region spanned by the vectors ${\vec r}_i^m$
no additional vector is required to delimit the region, as opposed to the
general case with non-orthogonal sides only bounding the region.  The
vectors ${\vec r}_i^m$ label points on the plane, and between any two
points which are adjacent in the ${\vec e}_1$ direction, a line is drawn
between them (adjacent means on the same $x_1$-axis).  Similarly, between
any two points adjacent in the ${\vec e}_2$ direction a line is drawn.
This integer $p$ then defines a bounded region in the plane, with general
disconnected components.  An alternative would be to fill in the entire 
rectangular region with points, and have the number parameterize all of 
the points.  The two representations are equivalent, but generate different 
numbers $p$.  

A set of further integers $p_j=a_1^{(j)} a_2^{(j)} \ldots a_n^{(j)}$
are used to label a stack of coplanar lattices with the same procedure
to fill in the third dimension.  The spacial filling of the disconnected
polhedron is assembled through the stacking of the base reduced
integers.

Colored polytopes are introduced by base reducing the integers $p_j$ into
the based reduced $a_j^{(k,m)}$ into base $N$.  The individual entries in
the lattice spanned by,

\bqr
{\vec r}= {\vec r}_i^m  = a_i^{m,k} \times m {\hat e}_1 +
       a_i^{m,k} \times i {\hat e}_2 + a_i^{m,k} \times k {\hat e}_3 \ .
\label{threedimpoly}
\fqr
The based reduced entries may be attributed into 'colors' or group
theory indices labeling a representation.

Next the volume $V$ and the $\partial V$ surface area of the polytope
region is deduced from the entries $a_i^{m,k}$.  The volume is the
sum of the individual entries $a_i^{m,k}$ over the entire lattice,

\bqr
p_j = a_i^k 2^i  \quad V_s= \sum_{i,k,m} a_i^{k,m} \ .
\fqr
The surface area of the polytope is a region bounded by the entries
of the entries $a_i^{m,k}$.  The bounded region is found via the
differences of the entries $a_i$; in two dimensions,

\bqr
V_{sf} = \sum_{ij} \vert a_i^{j} - a_{i-1}^{j}\vert -
  \sum_{ij} \vert a_i^{j} - a_i^{j-1} \vert \ .
\fqr
The region bounding the polytope is deduced from the differences
in the integers.

The terms in both series, $V_s$ and $V_{sf}$, are defined or
computed via the expansions,

\bqr
P_1^i = \sum M_{(1)}^{ij} p^j \qquad
P_2^i = M_{(2)}^{ij} p^j = \sum \vert a_i-a_{i-1}\vert_{{\rm p int}}
\ ,
\fqr
\bqr
P_1^i= \sum a_i\vert_{\rm p int} \ ,
\fqr
defined for the integer p configuration.
Even though the the individual terms $\vert a_i-a_i\vert$
in the summations involved the expansion are absolute value, the
entire sum is found via a summation over the individual numbers
$p$ parameterizing the lattice and its configuration.  (A
computation of Ising model partition function in one dimension
allows the matrices $M_1$ and $M_2$ to be computed indirectly for
particular lattices).

The 'colored' boundary is given a boundary via the same formalism,
but with a generalized difference $\vert a_i-a_{i-1}\vert$;
group theory or 'color' differences found with a different inner product
are possible.  The summations for these numbers may also be inverted
to obtain the values $a_i$ in terms of $p^j$ and an associated matrix.

An example list of this variables is given in the following table,

\bqr
\pmatrix{ p & a_i & p_1 & p_2 \cr 1 & 1 & 1 & 1 \cr
 2 & 01 & 1 & 2 \cr
 3 & 11 & 2 & 0 \cr
 4 & 001 & 1 & 2 \cr
 5 & 101 & 2 & 2 \cr
 6 & 011 & 2 & 2 \cr
 7 & 111 & 3 & 0 \cr } \ .
\label{partitions}
\fqr 

\noindent The number $p$ is listed, followed by the binary format; the integers
$p_1$ and $p_2$ are the sums $\sum a_i$ and $\sum \vert a_i - a_{i+1}\vert$,
in a cyclic fashion around the numbers $p$.

The polyhedron is constructed by the single numbers spanning the
multiple layers in 3-d, or by one number with the former grouped as
$p_1 p_2 \ldots p_n$.  The generalization to multiple dimensions is
straightforward.

The gluing of the polyhedra is clear.  For example, the numbers $p_1$
and $p_2$ that label two polytopes in two dimensions may be joined
by adding their base two reduced forms. The vectors ${\vec r}_1$ and
${\vec r}_2$ are added together to find ${\vec r}_3$; then ${\vec r}_3$
is modified to another number $p_3$.  If the overlap of the two initial
vectors results in a $2$ in the base two form, then there is intersection;
there should be an arithmetic operation on the two integers $p_1$ and
$p_2$ to find this answer.  For example, if $a_j{(1)}+a_j^{(2)}$ results
in a number greater than $2^n$ then there is overlap; this is for a
base $2^n$ number parameterizing a column of a square of dimension
$(n+1)\times (n+1)$.  The individual numbers $a_j$ in $p_1$ and $p_2$
add without overlap into the number $p$.

The rotations and translations of the individual polytopes may also be
formulated presumably as a functions operation on the number p.  These
operations have a direct application on the base $2$ form, by treating
the solids as a collection of vector points ${\vec r}$ and taking the
usual actions.  Changing the colors, when colored, is another operation.

These polytopic operations have many applications when the individual
numbers $p$ (i.e. the geometry) take on a dynamic setting, for example
in computing and cryptography, or when these numbers represent simplicial
complexes in a more physical application.

\section{Statistical Mechanics}

The summation over surfaces is required in statistical mechanics in order 
to compute the free energy and correlations.  This usually involves the 
summation of variables at large numbers of lattice sites.  The 
dimensionality of the lattice, and the couplings in these dimensions of 
the lattice points, complicates the solutions of these models.  The 
summation of variables at the lattice sites can be converted into 
a summation of random surfaces; the latter is made simpler by the 
polytopic definition and the conversion of the many body lattice integral 
into a discrete sum of integers that label all of the surfaces.  This is 
demonstrated in the following.  

The high temperature limit of the models is changed into a low temperature 
limit, in this formalism, via the the solution of the counting problem 
of the zero set to the level polynomials $P(z)=q$.   This is described 
in this section.  The mathematics is partially addressed in \cite{ChalmersZC}. 

Consider the $Z_N$ models defined by the Hamiltonian,

\bqr
H_0 = \sum \sigma_i \sigma_i \gamma_0^+
\fqr
\bqr
H_1 = \sum \sigma_i \sigma_{i\pm 1} \gamma_0^- + \sum \mu \sigma_i \ ,
\fqr
which in polytope language is,

\bqr
H = \sum_\Delta e^{-V_s\gamma_0^+ + (V-V_s)\gamma_0^- - V_{sf}4\gamma_1^+
 + 2 \partial V_{sf} \gamma_1^-} \ .
\fqr
The solids and surfaces count the $+$ and $-$ configurations, in
which the islands of $+$s and $-$s are polyhedra.  In the 2-d case, 
these models are typically solved, in a restricted coupling sense and 
for nearest neighbors, through transfer matrix methods.  An alternative
solution is via resolvants of the lattice configurations into the 
algebraic forms, i.e. integers, labeling them.

The partition function derived from these lattice configurations is,

\bqr
Z[\gamma_i,\mu] = \sum_{\sigma_i} e^{-\beta H} \ ,
\fqr
and via the partition sum, the free energy is,

\bqr
Z[\gamma_i,\mu] = \sum_p e^{-\beta H(p)} \ .
\label{partition}
\fqr
The pieces of the Hamiltonian are generated through the forms, 

\bqr
H_0 (p) = -(\gamma_0^+ - \gamma_0^-)  \sum M_{(1)}^{ij} p^j +
 ( N^d \gamma_0^- - 4\gamma_1^+)  \sum M_{(1)}^{ij} p^j
\fqr
\bqr
H_1(p) = \gamma_1^-  \sum M_{(2)}^{ij} p^j \ .
\fqr
and
\bqr
H_2(p) = -2 \mu  \sum M_{(1)} p^j + N^d \mu
\fqr
The polynomials in in $p$ label the volumes and surface areas of 
the polytopes involved in the sum.  In this approach the infinite 
number of variables is reduced to the single summation indexed by 
the polytope configuration variable $p$.  The polynomials $M_{i}^{ij} 
p^j$ are found in the previous section; the coefficients of the 
polynomials dictate the geometry of the lattice and also the form 
and number of interactions such as nearest neighbor and non-nearest 
neighbor.  A modification of the polynomials can incorporate all of 
these various interactions.    

The summations are expandable into,

\bqr
F(T;\gamma_i,\mu)= \sum_i \Delta(i) e^{-\gamma(i)}
\fqr
a partition into quasi-modular forms.  The function $\Delta(i)$ counts
the repetitions of the $\gamma(i)$ in the exponential expansion of the
partition function \rf{partition}.  Basically,

\bqr
\sum_{p} \prod_{i=1}^M e^{b_i p^i} = \sum N_p e^{-\gamma(p)} \ ,
\fqr
and the counting $N_p$ is found via solving for the zero set to 
the polynomials in the exponent of \rf{partition}.  These zeros are 
found from the solutions to

\bqr
\sum b_i p^i = \gamma(p) \ ,
\fqr
with $p$ an integer.  The countings of $p$ requires solving for
zeros of polynomial equations in one variable, with the degree
of the polynomial set by the number of lattice sites.  The form 
is derived from $M_{(i)}$, describing the interactions.

Consider the scenario of $\gamma_0^+=-\gamma_0^-=\gamma_1=\gamma$,
and no magnetic field.  Scaling the coupling constant out of the
partitions would generate an expansion in terms of $e^{m\beta\gamma}$.
The function $\Delta(i)$ is not coupling dependent, and the function
$\gamma(i)$ is $\beta\gamma m$.  The explicit coupling dependence
is
\bqr
\sum_{p} \prod_{i=1}^M e^{-\gamma\beta b_i p^i} =
  \sum N_n e^{-\beta \gamma(n)} \ ,
\fqr
with $\gamma(p)=\gamma p$.  The polynomial solutions to $\b_i p^i=n$
generate the high-temperature solution.  The polynomial nature also
makes the automorphicity somewhat apparent, because counting
the solutions have to be done.

The expansion generalizes to further interactions with,

\bqr
F(T;\gamma,i) = \sum_i \Delta_\gamma(i) \Delta_\rho(i)
 e^{-\gamma(i) - \rho(i)} \ .
\fqr
This occurs in the case of multiple interactions in the Hamiltonian,
for example, when the magnetic field is turned on.

Before closing this section, a few comments are made.  Most importantly,
the multiple summations on the spin variables $\sigma_i$, which is
large (near infinite), have been traded in for one variable, an integer.
This seems to be quite a simplification; however, the zeros of a
polynomial equation have to be performed (some progress along the lines 
in \cite{ChalmersZC} is required in order to make this explicit).  
The reduction of the system to one variable is quite important in the 
solution to these models in this approach.

Conversely, a solution to these models allows one to find zeros to
polynomials in special cases, e.g. model dependent.  The known solutions
for the models may be used, such as the Ising model in two dimensions.  
However, a more general statistical mechanical model is required to 
find more general zero level sets of polynomials.  

Multiple interactions may be included in these models.  For example,
the non-nearest neighbor interaction

\bqr
H_{n.n.} = \sum_i \gamma_s \sigma_{i} \sigma_{i\pm s}
\label{nonnearest}
\fqr
may be incorporated into the model, with the $i\pm s$ meaning that
there are interactions spaced a distance $s$ apart in specified
directions.  The complication in including these interactions is
in changing the polynomial equation $b_i p^i$.  The interactions
$\vert a_i - a_{i+s}\vert$ require new matrices $M_{(i)}$ which
change the $b_i$ matrices.  The matrix equations
$p^{(i)}_k = M_{(k)}^{ij} p^j$ require the $M_{(k)}$ to be computed.
Of course, the end result for the partition function with the
additional interactions requires only one sum to be computed.

The matrices $M_{(i)}$ are not included here for these interactions; 
however, they may be found for various lattices.  The solution of  
two-dimensional models is useful in the derivation for more general 
models.  Potentially all of the non-nearest neighbor interactions may be 
included, when the $M$ matrices are found that represent these couplings.  

Models with non-integer spin degrees of freedom may be examined, such as 
rational $p/q$ ones.  Also, perturbations with defect singularities, such 
as required with the Hubbard, can be placed in the models by changing the 
matrices $M_{(i)}$ appropriate to the defect couplings. 

The upshot of the analysis in the solution to the free energy of these 
models is that the two mathematical steps are required to be completed:  

(1) derivation of the appropriate matrices $M_{(i)}$ 

(2) derivation of the count to level sets of polynomials $P(z)=q$ 

\noindent These two steps are mathematically well-posed.  Their solution 
is important to solve most statistical mechanical models, and to uncover 
the structure beneath them.

\section{Numbers and the Polytopes to the Knots}

First a brief review of the definition and construction of a
polynomial invariant that uniquely characterizes the topology of
a contour in three dimensions is given (i.e. a knot invariant).  

The contour of the knot is labeled by an oriented line that self-intersects 
(over and under) at a number of points.  The knot configuration is made 
mathematically precise by labeling all of the oriented self-intersections.  
These four types of oriented intersections are labeled by two by two matrices; 
the collection of the matrices is put into a polynomial form by collecting the 
information in a systematic fashion.  These matrices are,

\bqr
M_1 = \pmatrix{ 1 & 0 \cr 0 & 0 }
  \qquad M_2 = \pmatrix{ 0 & 1 \cr 0 & 0 }
\fqr
\bqr
M_3 = \pmatrix{ 0 & 0 \cr 1 & 0 }
  \qquad M_4 = \pmatrix{ 0 & 0 \cr 0 & 1 } \ .
\label{intersections} 
\fqr 
There are a total of $n$ intersections
in the knot configuration, which through a single closed contour
are passed through twice each in traversing the loop. These
matrices are assembled into a $2n$ by $2n$ matrix $M$ via block
form by inserting at position (i,j) the two by two matrix
associated with the (i,j) node along the contour; this fills up
all but the diagonal elements. The diagonal entries along (i,i)
are given an empty two by two matrix.   Following the
arrows along the contour, the lower triangular two by two matrices 
are the transpose of the upper triangular ones and the matrix satisfies 
$M=M^T$. ($M_1$ is the geometric transpose of $M_4$).  

This matrix is a member of Sp($2$n) and gives a projection onto
the adjoint representation, $M=\sum_i a_i T^i$. One could put minus 
signs in the upper triangular portion so that the final matrix satisfies 
$M=-M^T$ to make it belong to SO($2n$).  The Sp($2n$) (or SO($2n$)) 
generators could be given the standard form,

\bqr
(M_{ab})^{ij} = \delta_a^i \delta_b^j \pm \delta_a^j \delta_b^i \ .
\label{matrixelement}
\fqr

Via the projection $M=\sum a_i T^i$ a polynomial is made that labels 
the knot.  The coefficients $a_i$ are assembled into the form $P(z)$,

\bqr
P(z) = \sum_{i=1}^{2n} a_i z^i \ , 
\label{invariant}
\fqr
a polynomial in the parameter $z$.  

\begin{figure}
\begin{center}
\epsfxsize=12cm
\epsfysize=12cm
\epsfbox{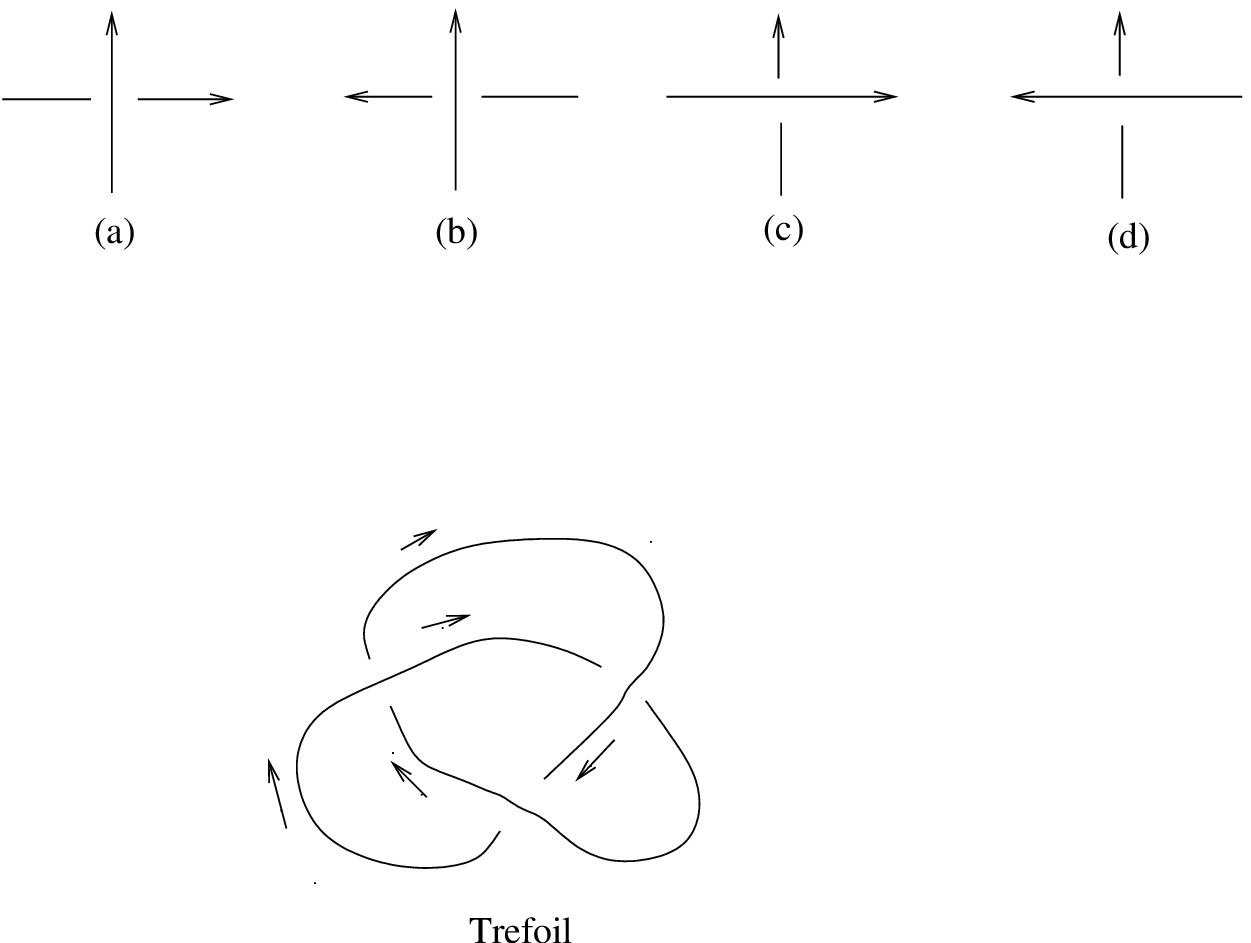}
\end{center}
\caption{(1) The four types of intersections.  (2) A sample
trefoil knot.}
\end{figure}

As an example, the trefoil knot's polynomial in figure 1 is given.  This 
configuration has three intersections and $M_t$ is dimension twelve.
The matrix $M_t$ in block form with the $M_j$ matrices
is,

\bqr
M_t = \pmatrix{ 0 & 0 & 0 & 2 & 0 & 0 \cr
                0 & 0 & 0 & 0 & 3 & 0 \cr
                0 & 0 & 0 & 0 & 0 & 2 \cr
                3 & 0 & 0 & 0 & 0 & 0 \cr
                0 & 2 & 0 & 0 & 0 & 0 \cr
                0 & 0 & 3 & 0 & 0 & 0 } \
\fqr
The decomposition of this trefoil's $M_t$ is $a_{8,1}=1$, $a_{9,4}=1$,
and $a_{12,5}=1$ and is symmetrized.  The polynomial $P_t(z)$ is, via
the decomposition of the generators through $z^{(j-1)*2n+i}$,

\bqr
P_t(z) = z^8 + z^{40} + z^{60} \ .
\fqr
This example describes the procedure for finding $M$ and $P(z)$.  

The knot polynomials described in \cite{ChalmersKnot} may be further 
reduced and put into the form of invariants, including the Reidemeister 
moves.  The polynomials have both number theory and group theory properties.  

Equivalent knots under the Reidemiester moves are grouped into polynomials 
of infinite degree, 

\bqr 
Q(z) = \sum b_i z^i \ ,
\fqr 
with the $b_i$ numbers representing the individual knot topologies.  Each 
$Q(z)$ represents an equivalence class of the topologies via the three 
Reidemeister moves, as discussed in \cite{ChalmersKnot}.  By definition, 
all of the $b_i$ are distinct numbers 
in not just the individual equivalence classes, but in all of the $Q(z)$ 
classes.  The number form of the knot invariant, as presented in 
\cite{ChalmersKnot}, is required; other knot invariants may also be 
placed in a number form.  

A use of the number theoretic knot form is that the geometries labeled by 
the polytopes may be mapped into the one-dimensional knot configurations 
embedded in three dimensions.  The use of such a mapping, basically from all 
topologies in more than one dimension to one-dimensional knot topologies 
is not entirely clear analytically, but can be used to classify topologies 
in higher dimensional geometries.  The reduction of the $d$-dimensions to 
$d=1$ might emphasizes the importance of knot mathematics and physics.  

The reduction to knot topologies of the higher dimensional topologies 
emphasizes further characterizations of the latter in terms of, for 
example, Reidemeister moves.  There could be unrecognized symmetries 
in the specifications of dimensionally varying topologies using the 
knot groups and knot characterizations.

Also, developments in transcendental computing based on number theoretic 
forms of higher dimensional geometries can be further classified and 
and possibly reduced in terms of these line elements and the symmetries 
inherit in them.  (Three-dimensional gauge physics and their correlations 
could be of use in this regard.)  

\section{Concluding remarks}  

Polytopes, i.e. polyhedra, within a volume V are mapped to integer numbers.  
The algebraic nature of the mapping of the topologies to the integers is 
relevant to descriptions of the topologies and their properties.  As an 
example, a function map of the cohomologies and homotopy based on the 
integers could be possible.  

The characterization of the multi-dimensional topologies can be mapped to 
one-dimensional knot configurations.  This property could lead to the 
manifestation of hidden symmetries in the topology, with relations 
to the former knot descriptions.  It seems possible that without the knot 
configurations a transcendental description of topology can be based in 
number theory and algebra also. 

The application to physics is clear especially in the field of statistical 
mechanics.  These models require means to sum over many variables at the 
the individual lattice sites and the polytopic definition can reduce this 
sum to a one variable summation.  Generalizations from nearest neighbor 
interactions to non-nearest neighbor interactions is clear from an algebraic 
standpoint.  

Required in the solution to the statistical mechanical models is that the 
two mathematical steps are required to be completed:  

(1) derivation of the appropriate matrices $M_{(i)}$ 

(2) derivation of the count to level sets of polynomials $P(z)=q$ 

\noindent These two steps are mathematically well-posed.  Their solution 
is important to solve most statistical mechanical models, and to uncover 
the structure beneath them.

The description of the polytopes in terms of integers has applications to 
math and physics.  There could be alternative descriptions of the surfaces 
that would lead to a simpler derivation of their uses; simpler means 
less computational transcendentally.  

The use in computational programming allows for an inherent parallel 
processing not described in this text based on geometric routing of 
data.  The integer description might be useful in this regard both in 
the assimilation and building of data streams.  Generalized RSA keys 
are also simple to construct based on fitting and placing of 
interlocking geometric solids.

\vfill\break

\end{document}